\documentclass[aps,prd,,twocolumn,groupedaddress,showpacs,nofootinbib]{revtex4}

\usepackage{graphicx,dcolumn,bm,amssymb,amsmath,latexsym,amsfonts,footnote}

\begin{document}

\newcommand{\nonu}{\nonumber}
\newcommand{\sm}{\small}
\newcommand{\noi}{\noindent}
\newcommand{\npg}{\newpage}
\newcommand{\nl}{\newline}
\newcommand{\bp}{\begin{picture}}
\newcommand{\ep}{\end{picture}}
\newcommand{\bc}{\begin{center}}
\newcommand{\ec}{\end{center}}
\newcommand{\be}{\begin{equation}}
\newcommand{\ee}{\end{equation}}
\newcommand{\beal}{\begin{align}}
\newcommand{\eeal}{\end{align}}
\newcommand{\bea}{\begin{eqnarray}}
\newcommand{\eea}{\end{eqnarray}}
\newcommand{\bnabla}{\mbox{\boldmath $\nabla$}}
\newcommand{\univec}{\textbf{a}}
\newcommand{\VectorA}{\textbf{A}}
\newcommand{\Pint}

\title{Unequal binary configurations of interacting Kerr-Newman black holes}

\author{I. Cabrera-Munguia\footnote{icabreramunguia@gmail.com}}
\affiliation{Departamento de F\'isica y Matem\'aticas, Universidad Aut\'onoma de Ciudad Ju\'arez, 32310 Ciudad Ju\'arez, Chihuahua, M\'exico}


\begin{abstract}
In this paper, binary systems of unequal Kerr-Newman black holes located on the axis and apart by a massless strut are investigated. After adopting a fitting parametrization, the conditions on the axis and the one eliminating the individual magnetic charges are solved exactly with the aim to obtain a $7$-parametric asymptotically flat exact solution. It is also deduced explicit analytical formulas for each half-length horizon $\sigma_{i}$, $i=1,2$, in terms of arbitrary physical Komar parameters: mass $M_{i}$, electric charge $Q_{i}$, angular momentum $J_{i}$, and a coordinate distance $R$. All the thermodynamical properties for the black holes as well as the interaction force related to the strut are also given in a concise way. Our analysis permits us to derive in detail the physical limits of the solution and introduce some scenarios where the strut is absent during the merger process.
\end{abstract}

\maketitle

\vspace{-0.4cm}
\section{Introduction}
\vspace{-0.3cm}
In the last years, the ongoing advances in the detection of gravitational waves produced by binary black hole (BH) mergers, has been leading us to develop analytical and exact models in order to complement the current powerful codes developed in numerical relativity. Although simplified binary BH models can be found in stationary spacetimes, these are very useful to conceive a general picture on how they might interact and evolve. In this direction, in an earlier paper \cite{Cabrera2018}, we were able to solve exactly the axis conditions with the main objective to describe dynamical scenarios between two unequal rotating Kerr BHs separated by a massless strut (conical singularity \cite{BachW,Israel}). These setups are well defined by means of an asymptotically flat metric that belongs to a $5$-parametric subclass of the well-known double-Kerr-NUT solution \cite{KramerNeugebauer}, but which is now fully characterized by arbitrary physical Komar parameters \cite{Komar}. This physical representation permits us to gain more novel information whether one attempts to study the dynamical aspects related to the spin-spin interaction (repulsion or attraction) within the binary system (BS), when the coalescence among two BHs is carrying out, or even before such a process could take place, and it can only be succeeded by getting concise expressions for the half-length BH horizons $\sigma_{1}$ and $\sigma_{2}$ in terms of Komar parameters. On the other hand, more general descriptions of stationary axisymmetric binary BHs can be analyzed once the electromagnetic field to the aforementioned rotating systems \cite{Cabrera2018} is added, nonetheless, it increases considerably the laborious task to obtain exact results, and for such a reason, alluding to their symmetric nature, we have considered first a pair of identical corotating charged binary BH models \cite{CCHV} where magnetic charges are not taken into account in the solution.

In the present paper we provide a $7$-parametric asymptotically flat exact solution that permits the description of the most general case of a BS composed of two unequal Kerr-Newman (KN) BHs \cite{ENewman} separated by a massless strut, where the functional form of their BH horizons as well as all the thermodynamical properties contained in the Smarr formula \cite{Smarr} are given by concise formulae. Due to the fact that the conditions on the axis and the absence of individual magnetic charges are fulfilled, the full metric is completely characterized by physical Komar parameters: the masses $M_{i}$, electric charges $Q_{i}$, and angular momenta $J_{i}$, while the centers of the sources are keeping apart by a coordinate distance $R$. In passing, we show that in a similar way as in the vacuum situation \cite{Cabrera2018}, the BHs interact via a dynamical law that is now defined by a septic algebraic equation, and when the strut is removed, it can be trivially reduced to some equilibrium laws already known; the case concerning vacuum systems \cite{MankoRuiz} and the electrostatic one \cite{Alekseev}. We also obtain quite trivial expressions for the thermodynamic relations of the BS during the merger limit (ML), whereas the result of Dietz and Hoenselaers \cite{DH} on the interaction force of two spinning particles at large distances has been generalized to include the contribution of the electric charges. Finally, in the seek of equilibrium states among the BHs without a supporting strut, some scenarios during the merger process have been found, where the new source that is being created might be seen as a BH, satisfying the subextreme condition $M^{2}-Q^{2}-(J/M)^{2}\geq0$ and avoiding the apparition of closed timelike curves (CTC) outside its event horizon \cite{Bonnor}.

The outline of the paper is as follows. In Sec. II we adopt a suitable parametrization for the double KN problem in order to get the corresponding asymptotically flat metric. Later on, the axis conditions and the one that eliminates both magnetic charges in the solution are solved exactly. In Sec. III, the thermodynamical features for each KN BH are given concisely once the expressions for the half-length BH horizons are derived. The dynamical limits of the interaction suffered by both BHs in the BS are well identified, in particular, we study equilibrium states with no strut during the merger process. Summary and outlook are introduced in Sec. IV.

\section{Suitable parametrization of the double KN problem}
\vspace{-0.4cm}
As has been previously outlined in Ref.\ \cite{CCHV}, the Ernst potentials $({\cal{E}}, \Phi)$ \cite{Ernst} on the upper part of the symmetry axis conveniently can be written in the form
\vspace{-0.1cm}
\begin{align}
{\cal E}(0,z)&=\frac{\mathfrak{e}_{1}}{\mathfrak{e}_{2}}, \qquad \Phi(0,z)=\frac{(Q+iB)z+\mathfrak{q}_{o}}{\mathfrak{e}_{2}}, \nonu \\
\mathfrak{e}_{1}&=z^{2}-[M + i(\mathfrak{q}+2J_{0})]z +\mathcal{P}_{+}+i P_{1} \nonu\\
&-2iJ_{0}  \bigg[M-i\mathfrak{q}+\frac{P_{2}}{\mathfrak{q}}\bigg], \nonu\\
\mathfrak{e}_{2}&=z^{2} + (M -i\mathfrak{q})z + \mathcal{P}_{-} + i P_{2}, \nonu\\
\mathcal{P}_{\pm}&= \frac{2\Delta_{o}-R^{2}-2 M\epsilon_{1}}{4} \mp \frac{\epsilon_{2}R-\mathfrak{q}{\rm S}_{1}+2(Q q_{o}+ B b_{o})}{2M}, \nonu\\
{\rm S}_{1}&=P_{1}+P_{2},\qquad \epsilon_{1,2}= \sigma_{1}^{2} \pm \sigma_{2}^{2}, \nonu\\
\Delta_{o}&= M^{2}-Q^{2}-B^{2}-\mathfrak{q}^{2}, \qquad
\mathfrak{q}_{o}=q_{o}+i b_{o},
\label{ernstaxiselectro}\end{align}

\vspace{-0.1cm}
\noi where the multipolar expansion \cite{Simon,HP,Sotiriou} enables us to show that $M$, $Q$, and $B$ are the total mass, total electric charge, and total magnetic charge of the binary setup, respectively. Moreover, the total electric and magnetic dipole moments, $Q_{o}$ and $B_{o}$, are given by the expressions
\vspace{-0.1cm}
\be Q_{o}=q_{o}-B(\mathfrak{q}+J_{0}), \qquad B_{o}=b_{o}+Q(\mathfrak{q}+J_{0}) \ee

\vspace{-0.1cm}
\noi while $J_{0}$ is the well-known NUT charge \cite{NUT} which is depicted as follows
\vspace{-0.1cm}
\begin{align} J_{0}&=\frac{\mathfrak{q}}{8M^{2}}
\left( \frac{N_{0}}{\mathfrak{q}^{2}P_{-}+P_{2}(P_{2}+M\mathfrak{q})} \right), \nonu \\
N_{0}&=M^{2}\left\{4(P_{1}P_{2}+|\mathfrak{q}_{o}|^{2})+(R^{2}-\Delta_{o})
(2\epsilon_{1}-\Delta_{o})+\epsilon_{2}^{2}\right\} \nonu \\
& -\left[\mathfrak{q}{\rm S}_{1}-\epsilon_{2}R-2(Q q_{o}+ B b_{o})\right]^{2}.
\label{NUTcharge}\end{align}

\vspace{-0.1cm}
For completeness, the total angular momentum of the system takes the simple aspect
\vspace{-0.1cm}
\be J=M \mathfrak{q}-\frac{{\rm S}_{2}}{2}+J_{0}\left(2M+\frac{P_{2}}{\mathfrak{q}}\right), \quad {\rm S}_{2}=P_{1}-P_{2}.\label{Multipolarterms}\ee

\vspace{-0.0cm}
\begin{figure}[ht]
\centering
\includegraphics[width=6.0cm,height=5.0cm]{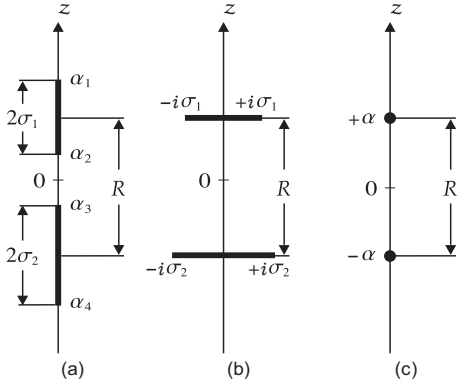}
\caption{Binary setups of unequal KN sources: (a) BH configuration when $\sigma_{i}^{2}>0$; (b) hyperextreme sources if $\sigma_{i} \rightarrow i \sigma_{i}$ (or $\sigma_{i}^{2}<0$ ); (c) extreme BHs for $\sigma_{i}=0$.}
\label{DK}\end{figure}

\vspace{-0.1cm}
The above Ernst potentials (the axis data) Eq.\ (\ref{ernstaxiselectro}) contain a total of eleven parameters within the set $\{M,\mathfrak{q},Q,B,q_{o},b_{o},R,\sigma_{1},\sigma_{2},P_{1},P_{2}\}$.\footnote{More particulars on the construction of the axis data Eq.\ (\ref{ernstaxiselectro})  can be found in Ref.\ \cite{CCHV}.} In addition these complex potentials fulfill the characteristic equation \cite{Sibgatullin,RMJ}
\vspace{-0.1cm}
\be {\cal E}(0,z) + \overline{{\cal E}}(0,z)+ 2 \Phi(0,z) \overline{\Phi}(0,z)=0,\label{characteristic}\ee

\vspace{-0.1cm}
\noi and its solution contains four roots $\alpha_{n}$ that determine the location of the sources. These are given by
\vspace{-0.1cm}
\be \alpha_{1,2}=\frac{R}{2}\pm \sigma_{1}, \qquad \alpha_{3,4}=-\frac{R}{2}\pm \sigma_{2},
\label{thealphas}\ee

\vspace{-0.1cm}
\noi where $R$ is a relative distance of separation between the sources, while the real or complex nature of the parameter $\sigma_{i}$ defines BHs if $\sigma^{2}_{i}\geq0$ or hyperextreme sources whether $\sigma_{i}^{2}<0$, as shown in Fig.\ \ref{DK}.

\vspace{-0.3cm}
\subsection{The asymptotically flat exact solution}
\vspace{-0.4cm}
With the aim to declare an asymptotically flat spacetime from the aforementioned axis data Eq.\ (\ref{ernstaxiselectro}), it is necessary to eliminate the gravitomagnetic monopole (or NUT charge). In order to simplify a little bit more the calculations of the next section, we are going to kill first the global magnetic charge $(B=0)$. So, the condition $J_{0}=0$ in Eq.\ (\ref{NUTcharge}) is achieved by virtue of
\vspace{-0.1cm}
\begin{align}
\epsilon_{1}&=\frac{\Delta}{2}\nonu\\
&+\frac{[\mathfrak{q}{\rm S}_{1}-\epsilon_{2} R-2Qq_{o}]^{2}-M^{2}\left[4(P_{1}P_{2}+|\mathfrak{q}_{o}|^2)+\epsilon_{2}^{2}\right]}{2M^{2}(R^{2}-\Delta)},\nonu\\
\Delta&= M^{2}-Q^{2}-\mathfrak{q}^{2}.\label{factorizes}\end{align}

\vspace{-0.1cm}
\noi and this result derives the axis data for asymptotically flat spacetimes, namely
\vspace{-0.1cm}
\begin{align}
{\cal E}(0,z)&=\frac{z^{2}-(M + i\mathfrak{q})z +P_{+}+i P_{1}}{z^{2} + (M -i\mathfrak{q})z + P_{-} + i P_{2}}, \nonu\\
\Phi(0,z)&=\frac{Qz+\mathfrak{q}_{o}}{z^{2} + (M -i\mathfrak{q})z + P_{-} + i P_{2}}, \nonu \\
P_{\pm}&= \frac{2\Delta-R^{2}-2 M\epsilon_{1}}{4} \mp \frac{\epsilon_{2}R-\mathfrak{q}{\rm S}_{1}+2Q q_{o}}{2M}. \label{ernstaxiselectroII}\end{align}

\vspace{-0.1cm}
Recalling that the complex potentials ${\cal{E}}=f-|\Phi|^{2}+i\Psi$ and $\Phi=-A_{4}+iA'_{3}$ satisfy the Ernst equations \cite{Ernst} for stationary axisymmetric systems in electrovacuum, which are given by
\vspace{-0.1cm}
\bea \begin{split}  \left({\rm{Re}} {\cal{E}}+|\Phi|^{2}\right)\Delta{\cal{E}}&=(\bnabla{\cal{E}}+
2\bar{\Phi}\bnabla \Phi)\cdot\bnabla {\cal{E}}, \\
\left({\rm{Re}}{\cal{E}}+|\Phi|^{2}\right)\Delta \Phi&=(\bnabla{\cal{E}}+
2\bar{\Phi}\bnabla\Phi)\cdot \bnabla\Phi, \label{ERNST} \end{split} \eea

\vspace{-0.1cm}
\noi while the metric functions $f$, $\omega$, and $\gamma$, define the Papapetrou line element \cite{Papapetrou}
\vspace{-0.1cm}
\be ds^{2}=f^{-1}\big[e^{2\gamma}(d\rho^{2}+dz^{2})+\rho^{2}d\varphi^{2}\big]- f(dt-\omega d\varphi)^{2}.
\label{Papapetrou}\ee

\vspace{-0.1cm}
A straight application of Sibgatullin's method \cite{Sibgatullin,RMJ} on Eq.\ (\ref{ernstaxiselectroII}) allows us to generate the Ernst potentials and metric functions in all the spacetime, where after some algebraic manipulations eventually it is possible to obtain
\vspace{-1.6cm}
\begin{widetext}
\begin{align}
{\cal{E}}&=\frac{\Lambda+\Gamma}{\Lambda-\Gamma},\qquad \Phi=\frac{\chi}{\Lambda-\Gamma}, \qquad f=\frac{|\Lambda|^{2}-|\Gamma|^{2}+ |\chi|^{2}}{|\Lambda-\Gamma|^{2}},\qquad \omega=2\mathfrak{q}+\frac{{\rm{Im}}\left[(\Lambda-\Gamma)(2z\overline{\Gamma}+ \overline{\mathcal{G}})-\chi \overline{\mathcal{I}} \right]}{|\Lambda|^{2}-|\Gamma|^{2}+ |\chi|^{2}},
\nonu\\
e^{2\gamma}&=\frac{|\Lambda|^{2}-|\Gamma|^{2}+ |\chi|^{2}}{256\sigma_{1}^{2}\sigma_{2}^{2}\kappa^{2} r_{1}r_{2}r_{3}r_{4}}, \qquad
\Lambda=4\sigma_{1}\sigma_{2} \left[\kappa(r_{1}+r_{2})(r_{3}+r_{4})+a(r_{1}-r_{3})(r_{2}-r_{4})\right]+\Big[2\kappa(\Delta-\epsilon_{1})-a R_{-}\Big]\nonu\\
&\times(r_{1}-r_{2})(r_{3}-r_{4})-16i\sigma_{1}\sigma_{2}
\Big \{ R\left[ \mathfrak{q} {\rm Re}(s_{1+})+ {\rm Im}(p_{1+}) \right]+(R^{2}+\epsilon_{2}){\rm Im}(s_{1+})-2\mathfrak{q}\sigma_{1}R_{1-} \Big\}r_{3}r_{4}\nonu\\
&+2i(\Lambda_{+}\mathfrak{r}_{1}-\Lambda_{-}\mathfrak{r}_{2}), \quad \Gamma=2\big(M\Gamma_{o}- b\chi_{+}\big), \quad
\chi=-2\big(Q\Gamma_{o}+2\mathcal{Q}\chi_{+}\big), \quad \Gamma_{o}=R \chi_{-}-2\sigma_{1}\sigma_{2}\chi_{s}+2\chi_{1+},\nonu\\
\Lambda_{\pm}&=\Big(\mathfrak{q} {\rm Re}(s_{1+})+ {\rm Im}(p_{1+})\Big)
\Big[ (R \pm c_{\mp})r_{3}-(R\pm c_{\pm})r_{4}\Big]+{\rm Im}(s_{1\pm})\big( R_{\pm}r_{3}-R_{\mp}r_{4} \big),  \nonu\\
\mathcal{G}&= 4\sigma_{1}\sigma_{2} \bigg\{ \Big[2R\big({\rm Re}(a)-2|\mathfrak{q}_{o}|^{2}\big)
+Q\big(QR(R^{2}-2\epsilon_{1})
+4q_{o}\epsilon_{2} \big)\Big](r_{1}r_{2}-r_{3}r_{4})
+i\big[2R{\rm Im}(a)+Q\xi-4\mathfrak{q}|\mathfrak{q}_{o}|^{2}\big]\nonu\\
&\times (r_{1}-r_{3})(r_{2}-r_{4})+4i\kappa(r_{2}r_{3}+r_{1}r_{4})\bigg\}
-2b(R\chi_{-}+2\sigma_{1}\sigma_{2}\chi_{s}),\nonu\\
&-R_{-}\bigg\{2c_{+}\Big[a-2\Big((R-c_{-}+i\mathfrak{q})s_{1-}+p_{1-} \Big)
(R-c_{+})\Big] +i\left(Q\xi+4Qb_{o}R_{+}
-4\mathfrak{q} |\mathfrak{q}_{o}|^{2}\right)\bigg\}(r_{1}-r_{2})(r_{3}-r_{4}) \nonu\\
&+\sigma_{2}\bigg\{4\Big[ 2\kappa(\Delta+\epsilon_{2})
-(R^{2}+\epsilon_{2}){\rm Re}(a)\Big]r_{4}+\Big[Q(R^{2}-\epsilon_{1})\xi_{1}-2Q\sigma_{1}^{2}\xi_{2}
+4(R^{2}+\epsilon_{2})|\mathfrak{q}_{o}|^{2}\Big](r_{3}+r_{4})\bigg\}(r_{1}-r_{2})\nonu\\
&+\sigma_{1}\bigg\{4\Big[ 2\kappa(\Delta-\epsilon_{2})-(R^{2}-\epsilon_{2}){\rm Re}(a)\Big]r_{2}
-\Big[Q(R^{2}-\epsilon_{1})\xi_{2}-2Q\sigma_{2}^{2}\xi_{1}-4(R^{2}-\epsilon_{2})|\mathfrak{q}_{o}|^{2}
\Big](r_{1}+r_{2})\bigg\}(r_{3}-r_{4}) \nonu\\
&+2M\Big[R^{2}\chi_{+}+2R\chi_{1-}+4\sigma_{1}\sigma_{2}(\chi_{p}-\chi_{2})\Big]
-2(Qb+2M\mathcal{Q})  \Big[\sigma_{2}(\overline{\kappa}_{1-}\mathfrak{r}_{1}
-\overline{\kappa}_{1+}\mathfrak{r}_{2})+ \sigma_{1}(\overline{\kappa}_{2+}\mathfrak{r}_{3}
-\overline{\kappa}_{2-}\mathfrak{r}_{4})\Big]\nonu\\
\mathcal{I}&=A\Big[4\sigma_{1}\sigma_{2}(r_{1}-r_{3})(r_{2}-r_{4})-R_{-}(r_{1}-r_{2})(r_{3}-r_{4})\Big]
+\kappa_{2+}\Big[B_{+}(R+c_{+})\mathfrak{r}_{1}-B_{-}(R-c_{-})\mathfrak{r}_{2} \Big]r_{4}\nonu\\
&-\kappa_{2-}\Big[B_{+}(R+c_{-})\mathfrak{r}_{1}-B_{-}(R-c_{+})\mathfrak{r}_{2} \Big]r_{3}-8\sigma_{1}\sigma_{2}\bigg\{\Big[ MR_{1-}\kappa_{1-}-B_{+}(2\mathfrak{q}_{o}R-\epsilon_{2}Q)\Big]r_{3}r_{4}-\kappa(Qb+2M\mathcal{Q})\bigg\}\nonu\\
&+ \Big[Q\big(3R^{2}-2\Delta-2\epsilon_{1}-2i{\rm S}_{1}\big)
+4i\mathfrak{q} \mathcal{Q}\Big]\chi_{+}+ 2(3R\mathcal{Q}-2\epsilon_{2}Q)\chi_{-}
+4\mathcal{Q}(\chi_{1+}+\sigma_{1}\sigma_{2}\chi_{s}) +4Q(R\chi_{1-}+2\sigma_{1}\sigma_{2}\chi_{p}), \nonu\\
\chi_{\pm}&=\sigma_{2}(s_{1+}\mathfrak{r}_{1}-s_{1-}\mathfrak{r}_{2}) \pm \sigma_{1}(s_{2+}
\mathfrak{r}_{3}-s_{2-}\mathfrak{r}_{4}), \quad
\chi_{1\pm}=\sigma_{2}(p_{1+}\mathfrak{r}_{1}+p_{1-}\mathfrak{r}_{2}) \pm \sigma_{1}(p_{2+}\mathfrak{r}_{3}+p_{2-}\mathfrak{r}_{4}),\nonu\\
\chi_{2}&=\sigma_{1}(s_{1+}\mathfrak{r}_{1}-s_{1-}\mathfrak{r}_{2}) + \sigma_{2}(s_{2+}\mathfrak{r}_{3}-s_{2-}
\mathfrak{r}_{4}),\quad
\chi_{p}=p_{1+}\mathfrak{r}_{1}-p_{1-}\mathfrak{r}_{2} + p_{2+}\mathfrak{r}_{3}-p_{2-}\mathfrak{r}_{4}, \quad \kappa=R_{+}R_{-},\nonu\\
\chi_{s}&=s_{1+}\mathfrak{r}_{1}+s_{1-}\mathfrak{r}_{2}+s_{2+}\mathfrak{r}_{3}+s_{2-}\mathfrak{r}_{4}, \quad
\xi=Q\Big[ 2M{\rm S}_{2}-\mathfrak{q}(R^{2}-2\Delta+2\epsilon_{1})\Big]-4\Big[ q_{o}{\rm S}_{1}+b_{o}(R^{2}-M^{2}+\mathfrak{q}^{2})\Big],\nonu\\
A&=M\Big[ \big[ 2 \mathcal{Q}+ Q(R-2\sigma_{1})\big]s_{1-}+2Qp_{1-}\Big]
+B_{-}\Big[Q\big( R^{2}-\Delta +\epsilon_{2}-i{\rm S}_{1}\big)-2(R+i\mathfrak{q})\mathfrak{q}_{o} \Big], \quad 
\mathcal{Q}=\mathfrak{q}_{o}+i \mathfrak{q} Q,\nonu\\
B_{\pm}&= [R s_{1\pm} \pm p_{1\pm} +Q \overline{\kappa}_{1\pm}]/M,  \quad p_{j\pm}=-2\sigma_{j}(R^{2}-\Delta)\pm i(-1)^{j}\Big[ M{\rm S}_{2}+2b_{o}Q+(R-i(-1)^{j}\mathfrak{q})
{\rm Im}(s_{j\pm})\Big],\nonu\\  
s_{j\pm}&=\Delta \mp 2\sigma_{j}R-(-1)^{j}\epsilon_{2}-i\Big[{\rm S}_{1}-\mathfrak{q}(R \mp 2\sigma_{j}) \Big],\quad
\kappa_{j \pm}=2\mathfrak{q}_{o}-(-i)^{j} R(Q\pm2\sigma_{j}), \quad \xi_{j}=4q_{o}R-(-1)^{j}Q(R^{2}-4\sigma_{j}^{2}), \nonu\\
a&=2(R+i\mathfrak{q})p_{1+}-s_{1+}\Big[\bar{s}_{2-}-2(R+i\mathfrak{q})(R-c_{-}+i\mathfrak{q})\Big],  \quad
b=\Big[ \mathfrak{q}{\rm S}_{1}-\epsilon_{2}R-2q_{o}Q+i M\big({\rm S}_{2}-2M\mathfrak{q}\big) \Big]/M, 
\nonu\\
R_{\pm}&=R^{2}-c_{\pm}^{2}, \quad c_{\pm}=\sigma_{1} \pm \sigma_{2},  \quad \mathfrak{r}_{1,2}=R_{1 \mp}r_{1,2}, \quad \mathfrak{r}_{3,4}=R_{2 \pm}r_{3,4},\quad R_{j\pm}=R^{2}\pm 2\sigma_{j}R-(-1)^{j}\epsilon_{2},\nonu\\
r_{1,2}&=\sqrt{\rho^{2}+(z-R/2 \mp \sigma_{1})^{2}},\quad 
r_{3,4}=\sqrt{\rho^{2}+(z+R/2 \mp \sigma_{2})^{2}}.
\label{Ernst9}\end{align}
\end{widetext}

\vspace{-0.3cm}
\subsection{Two unequal KN sources separated by a massless strut}
\vspace{-0.4cm}
If we are interested in the description of a BS composed by two KN sources; BHs or hyperextreme sources, therefore, we need to solve first the axis condition in the middle region of the symmetry axis, which reads as
\vspace{-0.1cm}
\be \omega\Big(\rho=0, {\rm{Re}}(\alpha_{3})<z< {\rm{Re}}(\alpha_{2})\Big)=0.\label{omegamiddle}\ee

\vspace{-0.3cm}
After placing Eq.\ (\ref{factorizes}) into this condition, eventually it is possible to obtain the following result
\vspace{-0.2cm}
\begin{widetext}
\begin{align}
&\mathfrak{q}(R+M)\big[2s_{o}+(R+M)Q^{2}\big]\epsilon_{2}^{2} +2\Big\{ \big[t_{o}s_{o} +\mathfrak{q}^{2}(R+M)(p_{o}-M^{2})\big]{\rm S}_{1} +2\mathfrak{q}q_{o}Q\big[s_{o}-(R+M)p_{o}\big]\Big\}\epsilon_{2}\nonu\\
&-\mathfrak{q}\Big\{2M^{2}(p_{o}{\rm S}_{2}^{2}-2Q^{2}P_{1}P_{2})-\big[ 2t_{o}(R^{2}+MR-\Delta)+\mathfrak{q}^{2}Q^{2}\big] {\rm S}_{1}^{2}\Big\}
+ 2M^{2}(R^{2}-\Delta)\big[M\mathfrak{q}^{2}+(R+M)(p_{o}+Q^{2})\big]{\rm S}_{2}\nonu\\
&+4q_{o}Q\big[(M^{2}+\mathfrak{q}^{2})p_{o}-t_{o}(R^{2}+MR-\Delta) \big]{\rm S}_{1}+4\mathfrak{q}q_{o}^{2} \big[M^{2}P_{0}-Q^{2}(2p_{o}+Q^{2})\big]+
4b_{o}M^{2}P_{0}\big[\mathfrak{q}b_{o}+Q(R^{2}-\Delta)\big]\nonu\\
&-M^{2}\mathfrak{q}(2MR+2M^{2}-Q^{2})(R^{2}-\Delta)^{2}=0,\nonu\\
P_{0}&=(R+M)^{2}+\mathfrak{q}^{2}, \quad p_{o}=R^{2}+MR+\mathfrak{q}^{2},\quad
s_{o}=M\mathfrak{q}^{2}-(R+M)(R^{2}-\Delta), \quad  t_{o}=M^{2}-\mathfrak{q}^{2}+MR.  \label{conditionmiddle}
\end{align}

Despite it has been previously assumed $B=0$ to remove the global magnetic charge, the solution describes itself two unequal sources endowed with identical magnetic charges but carrying opposite signs. The presence of the Dirac string connecting the sources is unavoidable unless both magnetic charges have been completely eliminated. This task may be accomplished when is solved the condition written below \cite{Tomi}
\vspace{-0.1cm}
\begin{align}
&{\rm Re}(\Phi)\Big(\rho=0,z={\rm{Re}}(\alpha_{2i-1})\Big)-{\rm Re}(\Phi)\Big(\rho=0,z={\rm{Re}}(\alpha_{2i})\Big)=0,\nonu\\
i&=1,2, \label{nomagneticcharge}\end{align}

\vspace{-0.1cm}
\noi and taking into account once again Eq.\ (\ref{factorizes}), the condition killing both magnetic charges and, therefore, the Dirac string, is given by
\vspace{-0.1cm}
\begin{align}
&\mathfrak{q} Q(R+M)^{3}\epsilon_{2}^{2}+(R+M)\Big\{ Q\big[MP_{0}-2\mathfrak{q}^{2}(R+M)\big]
{\rm S}_{1}-2\mathfrak{q}q_{o}\big[MP_{0}-2(R+M)Q^{2} \big]\Big\}\epsilon_{2}\nonu\\
&+\mathfrak{q}Q \Big\{ \big[ MP_{0}-(R+M)(2M^{2}+2MR-\mathfrak{q}^{2})\big]{\rm S}_{1}^{2}-4M^{2}(R+M)
P_{1}P_{2}\Big\}+M^{2}\big[2\mathfrak{q}P_{0}b_{o}-Q\left(P_{0}-2\mathfrak{q}^{2}\right)
(R^{2}-\Delta) \big]{\rm S}_{2}\nonu\\
&-2q_{o}\big[ M(\Delta+M R)P_{0}+2\mathfrak{q}^{2}(R+2M)Q^{2}\big]{\rm S}_{1} -4\mathfrak{q}q_{o}^{2}Q\big[ MP_{0}-(R+M)Q^{2}\big]\nonu\\
&-M^{2}(M\mathfrak{q}^{2}-s_{o}) \big[ 2P_{0}b_{o}+\mathfrak{q} Q(R^{2}-\Delta) \big]=0.  \label{conditionnoBcharge}
\end{align}
\end{widetext}

\vspace{-0.1cm}
Note that both Eqs.\ (\ref{conditionmiddle}) and (\ref{conditionnoBcharge}) represent a pair of quadratic algebraic equations defined by any of the variables $\epsilon_{2}$, $q_{o}$, $P_{1}$, and $P_{2}$. The easiest solution satisfying Eqs.\ (\ref{conditionmiddle}) and (\ref{conditionnoBcharge}) appears when $\mathfrak{q}=0$, leading us to
\vspace{-0.2cm}
\be P_{1,2}=-\frac{b_{o}}{M}\Bigg(\frac{(R^{2}-M^{2}+Q^{2})(MR\pm\epsilon_{2}),}{2q_{o}R-\epsilon_{2}Q}\pm Q\Bigg),\ee

\vspace{-0.1cm}
\noi where the signs $+$ and $-$ are associated to the subscripts $1$ and $2$. In this respect, it represents two unequal counterrotating KN BHs that are apart by a massless strut, which has been considered before in Ref.\ \cite{ICM}. However, regardless of such a trivial case, these two quadratic equations are solved \emph{exactly} in the most general case through the following parametrization:
\vspace{-0.1cm}
\begin{widetext}
\begin{align}
\epsilon_{2}&=- \frac{t_{o}{\rm S}_{1}+Mr(R^{2}-\Delta)+2\mathfrak{q} q_{o}Q}{\mathfrak{q}(R+M)},\quad
q_{o}=\frac{(M\mathfrak{q}^{2}-s_{o})(\mathfrak{q} \delta_{1}-Qr)-
QP_{0}{\rm S}_{1}-\mathfrak{q}^{2}\delta_{1}{\rm S}_{2}}{2\mathfrak{q}P_{0}},\nonu \\
b_{o}&= \frac{(R^{2}-\Delta)(\delta_{1} r-\mathfrak{q} Q)-Q(R+M){\rm S}_{2}}{2P_{0}}, \nonu\\
P_{1,2}&=\Bigg(\frac{s_{o}}{2P_{0}}-\frac{\mathfrak{q}^{2}(R+M)Q^{2}}{4P_{0}^{2}}\Bigg)r
-\frac{\mathfrak{q}^{2}(R+M)}{(R^{2}-\Delta)^{2}} 
\Bigg(\frac{2p_{o}}{P_{0}}+ \frac{\mathfrak{q}^{2}(Q^{2}-\delta_{1}^{2})}{P_{0}^{2}} \Bigg)\frac{J^{2}}{r} + \Bigg(\frac{\mathfrak{q}^{2}Q\delta_{1}}{P_{0}(R^{2}-\Delta)} \mp 1 \Bigg)J\nonu\\
& + \mathfrak{q}\Bigg(M \mp \frac{(M\mathfrak{q}^{2}+s_{o})Q\delta_{1}}{2P_{0}(R^{2}-\Delta)} \Bigg)
+\frac{\mathfrak{q}(R+M)}{(R^{2}-\Delta)^{2}}\Bigg[2M\mathfrak{q}^{2}\Bigg(1-\frac{Q^{2}}
{P_{0}}\Bigg) + \frac{\mathfrak{q}^{2}(M\mathfrak{q}^{2}+s_{o})(Q^{2}-\delta_{1}^{2})}{P_{0}^{2}}
-R(R^{2}-\Delta)\Bigg] \frac{J}{r}\nonu \\
&-\frac{\mathfrak{q}^{2}(R+M)(M\mathfrak{q}^{2}+s_{o})}{4(R^{2}-\Delta)} \Bigg[2M\bigg(1-\frac{Q^{2}}{P_{0}}\bigg) + \frac{(M\mathfrak{q}^{2}+s_{o})(Q^{2}-\delta_{1}^{2})}{P_{0}^{2}} \Bigg]\frac{1}{r}.
\label{solution}\end{align}
\end{widetext}

\vspace{-0.1cm}
\noi Due to the fact that in the BS one can interchange the location of its components as well as their physical properties, it is worth noting that our analytical result solving exactly Eqs.\ (\ref{conditionmiddle}) and (\ref{conditionnoBcharge}), accomplishes the conditions $\{P_{1,2},q_{o},b_{o},\epsilon_{1},\epsilon_{2}\} \rightarrow \{-P_{2,1},-q_{o},b_{o},\epsilon_{1},-\epsilon_{2}\}$, under the change $\{J,r\}\rightarrow \{J,-r\}$. Moreover, in the absence of electromagnetic field; i.e., when $Q=0$, $q_{o}=0$, $b_{o}=0$, and
\vspace{-0.1cm}
\begin{align} J&=M\mathfrak{q}+\frac{(R^{2}-M^{2}+\mathfrak{q}^{2})(\mathfrak{q}-a_{1}-a_{2})}{2(R+M)}, \nonu\\ r&=a_{1}-a_{2},\label{conditionvacuum}\end{align}

\vspace{-0.1cm}
\noi the above exact solution is reduced to the one earlier studied in Ref.\ \cite{Cabrera2018} for vacuum systems, where $a_{i} \equiv J_{i}/M_{i}$ defines the Komar angular momentum per unit mass of each source.

At this point, we would like to remark that our analytical result solving the axis conditions in combination with the absence of individual magnetic charges is quite convenient in the description of BHs and hyperextreme sources (or naked singularities). This statement is supported by the fact that our ansatz satisfying these physical requirements involves the squares of the quantities $\sigma_{1}$ and $\sigma_{2}$ in terms of the remaining parameters. Nonetheless, in the remainder of this paper, we will focus our attention to the BH sector.

\vspace{-0.3cm}
\section{Thermodynamical and dynamical aspects of the double-KN solution}
\vspace{-0.4cm}
Because $\sigma_{i}^{2}\geq0$ defines a BH, a more transparent and physical representation of the double-KN solution might be achieved if we are able to entirely express the BH horizons in terms of Komar's parameters. To perform such a task we have at hand the well-known Tomimatsu's formulas for stationary axisymmetric spacetimes in electrovacuum \cite{Tomi,Galtsov} \footnote{The absence of magnetic charges in the BS avoids the contribution of the term $\frac{1}{8\pi}\int_{H_{i}} (A_{3}^{'}A_{3})_{,z}\, d\varphi dz$ in the above Eq.\ (\ref{Tomy}), preserving the conventional Smarr formula for the mass.},
\vspace{-0.1cm}
\begin{align} M_{i}&= -\frac{1}{8\pi}\int_{H_{i}} \big[\omega \Psi_{,z}-2(A_{3}^{'}A_{3})_{,z}\big]\, d\varphi dz, \nonu\\
Q_{i}&=\frac{1}{4\pi}\int_{H_{i}}\omega A_{3,z}^{'}\, d\varphi dz, \nonu\\
J_{i}&=-\frac{1}{8\pi}\int_{H_{i}}\omega\left[1+ \frac{\omega \Psi_{,z}}{2}
-\tilde{A}_{3}A_{3,z}^{'}-(A_{3}^{'}A_{3})_{,z}\right]d\varphi dz, \label{Tomy} \end{align}

\vspace{-0.1cm}
\noi with $\tilde{A}_{3}=A_{3}+ \omega A_{4}$ and $\Psi={\rm Im}(\cal{E})$, where the magnetic potential $A_{3}$ is the real part of Kinnersley's potential $\Phi_{2}$ \cite{Kinnersley}, thus having
\vspace{-0.1cm}
\begin{align} A_{3}&={\rm Re}(\Phi_{2})=-2\mathfrak{q} A_{4}-zA'_{3}+{\rm Im} \bigg(\frac{\mathcal{I}}{\Lambda-\Gamma}\bigg), \nonu\\
\Phi_{2}&=\frac{(2\mathfrak{q}+iz)\chi-i\mathcal{I}}{\Lambda-\Gamma}, \label{Kinnersley} \end{align}

\vspace{-0.1cm}
\noi where we must bear in mind that $-A_{4}$ and $A'_{3}$ are the real and imaginary components of the Ernst potential $\Phi$, respectively. The integrals shall be evaluated over the BH horizons, which in cylindrical coordinates $(\rho,z)$ are looked as thin rods having a domain inside the intervals $\alpha_{2i}\leq z \leq \alpha_{2i-1}$, \, $i=1,2$. A combination of Eqs.\ (\ref{Tomy}) allows us to prove that each BH satisfies the Smarr formula \cite{Smarr,Tomi}
\vspace{-0.1cm}
\begin{align}
M_{i}&=\sigma_{i}+2\Omega_{i}J_{i}+ \Phi_{i}^{H}Q_{i}=\frac{\kappa_{i}S_{i}}{4\pi}+2\Omega_{i}J_{i}+\Phi_{i}^{H}Q_{i},\nonu\\
i&=1,2. \label{Massformula}\end{align}

\vspace{-0.1cm}
\noi On one hand $\Omega_{i}$ and $\Phi_{i}^{H}$ are the angular velocity and the electric potential in the rotating frame of the $i \rm {th}$ BH horizon, respectively. On the other hand, $S_{i}$ is the area of the horizon while $\kappa_{i}$ represents the surface gravity. Before deriving the thermodynamical properties contained in the Smarr formula, it is necessary to get first both analytical expressions of the BH horizons as a function of physical Komar parameters. In order to carry out this rather involved assignment, one substitutes Eqs.\ (\ref{factorizes}), (\ref{Ernst9}), (\ref{solution}), and (\ref{Kinnersley}), inside Eq.\ (\ref{Tomy}), to obtain the corresponding masses $M_{i}$ and electric charges $Q_{i}$ for both BHs. The result is
\vspace{-0.0cm}
\begin{align}
M_{1,2}&=\frac{M\pm \delta_{2}}{2},\qquad Q_{1,2}=\frac{Q\pm \delta_{1}}{2},\nonu \\
\delta_{2}&=\frac{(J-M\mathfrak{q})\left[2P_{0}(R^{2}+MR+\mathfrak{q}^{2})+\mathfrak{q}^{2}(Q^{2}-\delta_{1}^{2}) \right]}{r P_{0}(R^{2}-\Delta)}\nonu \\
&+\frac{\mathfrak{q}\left[2MP_{0}+(Q^{2}-\delta_{1}^{2})(R+M)\right]}
{2rP_{0}},\label{massesandcharges}\end{align}

\vspace{-0.1cm}
\noi and it can be noticed that the total mass $M=M_{1}+M_{2}$ and total electric charge $Q=Q_{1}+Q_{2}$, whereas $\delta_{2}=M_{1}-M_{2}$ and $\delta_{1}=Q_{1}-Q_{2}$. In addition, the two angular momenta are given by
\vspace{-0.1cm}
\begin{align}
J_{1}&= \frac{\mathfrak{q}(R^{2}-\Delta)H_{1+}H_{1-}+(J-M\mathfrak{q})\mathcal{P}_{1}}
{P_{0}^{2}(R^{2}-\Delta)(M_{1}-M_{2})}, \nonu\\ J_{2}&=\frac{\mathfrak{q}(R^{2}-\Delta)H_{2+}H_{2-}
+(J-M\mathfrak{q})\mathcal{P}_{2}}
{P_{0}^{2}(R^{2}-\Delta)(M_{2}-M_{1})},\nonu\\
\mathcal{P}_{i}&=H_{i-}C_{i}-(-1)^{i}(M_{1}-M_{2})Q^{2}_{i}P_{0}^{2},\nonu\\
C_{i}&=P_{0}^{2}-2M_{i}(R+M)P_{0}+2\mathfrak{q}^{2} Q_{1}Q_{2},\nonu\\
H_{i\pm}&=M_{i}P_{0}\pm Q_{1}Q_{2}(R+M).
\label{angularmomenta}\end{align}

\vspace{-0.1cm}
\noi whose sum is exactly the total angular momentum of the BS; i.e., $J=J_{1}+J_{2}$. In a similar way like in the vacuum case, the appropriate dynamical law for interacting KN sources with struts can be accomplished via trivial algebraic manipulations on both expressions contained in Eq.\ (\ref{angularmomenta}). This law has the simplified aspect
\vspace{-0.1cm}
\begin{align}
&\mathfrak{q}\mathcal{P}_{0}-J_{1}\mathcal{P}_{2}-J_{2}\mathcal{P}_{1}=0, \nonu\\
&\mathcal{P}_{0}=M\mathcal{P}_{1}-(R^{2}-\Delta)H_{1+}H_{1-}\nonu\\
&\equiv M\mathcal{P}_{2}-(R^{2}-\Delta)H_{2+}H_{2-}.
\label{condition}\end{align}

\vspace{-0.1cm}
After combining the above set of Eqs.\ (\ref{massesandcharges}) and (\ref{angularmomenta}) it can be doable to get two relations that completely characterize the exact result given by Eq.\ (\ref{solution}) in terms of arbitrary physical Komar parameters, namely
\vspace{-0.1cm}
\begin{align}
J&=M \mathfrak{q}+\frac{2(R^{2}-\Delta)N_{1}}{\mathcal{P}_{1}+\mathcal{P}_{2}-MP_{0}^{2}(R^{2}-\Delta)},  \nonu\\
r&=\frac{2P_{0}N_{2}}{M(\mathcal{P}_{1}+\mathcal{P}_{2})-(M_{1}-M_{2})^{2}P_{0}^{2}(R^{2}-\Delta)},\nonu\\
N_{1}&=M_{1}M_{2}P_{0}^{2}(\mathfrak{q}-a_{1}-a_{2})+\mathfrak{q}Q_{1}^{2}
Q_{2}^{2}(R+M)^{2},\nonu\\
N_{2}&= M_{1}M_{2} (C_{1}+C_{2})(a_{1}-a_{2})-\mathfrak{q}\Big[Q(M_{2}Q_{1}-M_{1}Q_{2})\nonu\\
&\times (H_{1+}+H_{2+})-(M_{1}-M_{2})Q_{1}Q_{2}RP_{0}\Big],
\label{conditionparameters}\end{align}

\vspace{-0.1cm}
\noi where it is observed that after placing $Q_{1}=Q_{2}=0$ into Eq.\ (\ref{conditionparameters}), it brings us immediately the aforementioned Eq.\ (\ref{conditionvacuum}), which describes vacuum systems \cite{Cabrera2018}. Moreover, by using these two expressions contained above for $J$ and $r$, the algebraic quantities $P_{1}$ and $P_{2}$ can be written down in a more compact form
\vspace{-0.1cm}
\begin{align}
P_{1,2}&=\frac{(2H_{2}A_{2}-RP_{0}\mathcal{P}_{2})J_{1}-(2H_{1}A_{1}-RP_{0}\mathcal{P}_{1})J_{2}}{2P_{0}\mathcal{P}_{0}}
\nonu\\ 
&\pm (M\mathfrak{q}-J),\nonu\\
A_{i}&=\mathcal{P}_{i}-(R^{2}-\Delta)H_{i+}P_{0}, \nonu\\ 
H_{i}&=M_{i}P_{0}-Q_{i}Q(R+M),
\label{simplePx}\end{align}

\vspace{-0.1cm}
\noi while the electric and magnetic dipole moments are reduced to
\vspace{-0.1cm}
\begin{widetext}
\begin{align}
Q_{o}&=\frac{2\mathfrak{q} (Q_{1}J_{2}-Q_{2}J_{1})}{P_{0}}+\frac{1}{2} Q_{1}(R-2M_{2})-\frac{1}{2} Q_{2}(R-2M_{1}),\nonu\\
B_{o}&=\mathfrak{q} Q+\Bigg[(Q_{1}C_{2}-Q_{2}C_{1})\bigg(\frac{J_{1}}{\mathcal{P}_{1}}-\frac{J_{2}}{\mathcal{P}_{2}}\bigg)
-\frac{\mathfrak{q}}{P_{0}}\Bigg(Q -\frac{Q_{1}H_{1+}\big[Q_{1}(Q_{1}-Q_{2})P_{0}-2(R+M)H_{1-}\big]}
{\mathcal{P}_{1}} \nonu\\
&-\frac{Q_{2}H_{2+}\big[Q_{2}(Q_{2}-Q_{1})P_{0}-2(R+M)H_{2-}\big]}{\mathcal{P}_{2}} \Bigg) \Bigg] \frac{(R^{2}-\Delta)}{2}.
\label{simpleQx}\end{align}
\end{widetext}

\vspace{-0.1cm}
Finally, as a result of combining Eqs.\ (\ref{factorizes}), (\ref{solution}), (\ref{condition}), and (\ref{conditionparameters}), together with a few non-trivial efforts, the explicit formula for both unequal BH horizons $\sigma_{i}$ in a physical representation are
\vspace{-0.1cm}
\begin{align}
\sigma_{i}&=\sqrt{D_{i}-J_{i}\left( \frac{J_{i}G_{i}-2\mathfrak{q}A_{i}B_{i}}{P_{0}^{2}\mathcal{P}_{i}^{2}}\right)},\nonu\\
D_{i}&=M_{i}^{2}- Q_{i}^{2}F_{i}-2(-1)^{i}Q_{i}F_{0}, \nonu\\
G_{i}&=\left[2(R+M)\mathcal{P}_{i}+P_{0}(R^{2}-\Delta)C_{i}\right]^{2}-4P_{0}\mathcal{P}_{1}\mathcal{P}_{2},\nonu\\
F_{0}&=\frac{M_{2}Q_{1}-M_{1}Q_{2}}{R+M} \bigg( 1-\frac{\mathfrak{q}^{2}}{P_{0}}\bigg),
 \nonu\\  F_{i}&=1-\frac{Q_{i}^{2}\mathfrak{q}^{2}}{P_{0}^{2}}
\bigg(1-\frac{A_{i}^{2}}{\mathcal{P}_{i}^{2}}\bigg)+\frac{Q^{2}\mathfrak{q}^{2}}{P_{0}^{2}},\nonu\\
B_{i}&=Q_{i}^{2}P_{0}(R^{2}-\Delta)C_{i}-2H_{i}\mathcal{P}_{i}, \quad i=1,2.\label{sigmas}\end{align}

\vspace{-0.1cm}
We observe that the above expressions contained in Eq.\ (\ref{sigmas}) acquire a symmetric character since both horizons can be obtained from each other under the change of their constituents; i.e., $\sigma_{2}=\sigma_{1(1\leftrightarrow2)}$. In the absence of electromagnetic field, the horizons adopt the simplified form \cite{Cabrera2018}
\vspace{-0.1cm}
\begin{align}
\sigma_{1}&=\sqrt{M_{1}^{2}-a_{1}^{2}+\gamma_{12}},\qquad \sigma_{2}=\sqrt{M_{2}^{2}-a_{2}^{2}+\gamma_{21}},\nonu\\
\gamma_{12}&=4a_{1}M_{2}\frac{a_{1}M_{2}\mathfrak{q}^{2}+
[M_{1}(\mathfrak{q}+a_{1}-a_{2})+a_{1}R]P_{0}}
{P_{0}^{2}},\nonu\\
\gamma_{21}&=4a_{2}M_{1}\frac{a_{2}M_{1}\mathfrak{q}^{2}+
[M_{2}(\mathfrak{q}-a_{1}+a_{2})+a_{2}R]P_{0}}
{P_{0}^{2}},
 \label{sigmasvacuum}\end{align}

\vspace{-0.1cm}
\noi where the physical parameters $\{M_{1},M_{2},J_{i},J_{2},R\}$ are related to each other via a cubic equation
\vspace{-0.1cm}
\begin{align}
&\mathfrak{q} P_{0}-a_{1}p_{1}-a_{2}p_{2}=0, \nonu\\
p_{1}&=(R+M_{1})^{2}-M_{2}^{2}+\mathfrak{q}^{2}, \nonu\\
p_{2}&=(R+M_{2})^{2}-M_{1}^{2}+\mathfrak{q}^{2}.
\label{conditionvacuumII}\end{align}

\vspace{-0.1cm}
The reader should be aware that this last equation is nothing less than Eq.\ (\ref{conditionvacuum}), which has been differently written; it represents the non-electrically charged version of Eq.\ (\ref{condition}). Actually, it should be pointed out that unlike the vacuum case, the dynamical law Eq.\ (\ref{condition}) does not give us the opportunity to express the BH horizons in terms of the seven physical parameters $\{M_{1},M_{2},Q_{1},Q_{2},J_{i},J_{2},R\}$ in a explicit form, because it corresponds to a seventh-degree equation in terms of the variable $\mathfrak{q}$, therefore, one must make use of numerical analysis. Furthermore, in the lack of rotation; i.e., $\mathfrak{q}=0$ and $J_{1}=J_{2}=0$, there exists no dynamical law and the horizons become electrostatics; both are given by \cite{VCH}
\vspace{-0.1cm}
\begin{align} \sigma_{i}&=\sqrt{M_{i}^{2}- Q_{i}^{2}-2(-1)^{i}Q_{i}\frac{M_{2}Q_{1}-M_{1}Q_{2}}{R+M_{1}+M_{2}}}, \nonu\\
i&=1,2. \label{VarzChist} \end{align}

\vspace{-0.1cm}
Once we already know the functional form of $\sigma_{1}$ and $\sigma_{2}$, our experience acquired in the vacuum scenario \cite{Cabrera2018}, suggests us that it might be possible to determine simple expressions for the thermodynamical characteristics of the BS displayed in the Smarr formula Eq.\ (\ref{Massformula}), indeed this will be the case, as we shall see next. First of all, we have that the area of the horizon and surface gravity can be computed via the formulas \cite{Tomi,Carter}
\vspace{-0.1cm}
\be S_{i}=\frac{4\pi \sigma_{i}}{\kappa_{i}}, \qquad \kappa_{i}= \sqrt{-\Omega_{i}^{2}e^{-2\gamma^{H_{i}}}},\ee

\vspace{-0.1cm}
\noi where $\Omega_{i}= \omega_{i}^{-1}$. Reminding that $\omega_{i}$ and $\gamma^{H_{i}}$ are the constant values that metric functions $\omega$ and $\gamma$ take on the axis part denoting the horizon $H_{i}$. Secondly, the electric potential $\Phi_{i}^{H}$ can be obtained in a straightforward manner by using the Smarr formula Eq.\ (\ref{Massformula}). Cumbersome calculations permits us to determine the following concise results
\vspace{-0.1cm}
\begin{align}
\Omega_{i}&= \frac{\mathfrak{q} A_{i}}{P_{0}\mathcal{P}_{i}} + \frac{J_{i}P_{0}^{3}\mathcal{P}_{i}(R^{2}-\Delta)
\Big[(R+\sigma_{i})^{2}-\sigma_{j}^{2}\Big]}{\mathcal{P}_{i}^{2}\mathcal{N}_{i}^{2}+
P_{0}^{2}(R^{2}-\Delta)^{2}\mathcal{M}_{i}^{2}}, \nonu\\
\frac{S_{i}}{4\pi}&= \frac{\mathcal{P}_{i}^{2}\mathcal{N}_{i}^{2}+
P_{0}^{2}(R^{2}-\Delta)^{2}\mathcal{M}_{i}^{2}}{P_{0}\mathcal{P}_{i}^{2}
\left[(R+\sigma_{i})^{2}-\sigma_{j}^{2}\right]}, \nonu\\
\Phi_{i}^{H}&= \frac{M_{i}-\sigma_{i}-2\Omega_{i}J_{i}}{Q_{i}},\nonu\\
\mathcal{N}_{i}&=P_{0}(M_{i}+\sigma_{i})-2\mathfrak{q}J_{i}-Q_{i}Q(R+M),\nonu\\
\mathcal{M}_{i}&=J_{i}C_{i}+\mathfrak{q}Q_{i}^{2}H_{i+},\qquad
i,j=1,2, \quad i \neq j. \label{Horizonproperties}\end{align}

\vspace{-0.1cm}
Then we have that for a binary KN BH configuration, the peculiar aspect of the formula for angular velocity reveals a natural mechanism to better interpret the induced angular velocity in a source that has lost its own angular momentum. For instance, if we assume that this is case for the second BH, then $J_{2}=0$, and its non zero angular velocity is
\vspace{-0.1cm}
\begin{align} \Omega_{2}&=\frac{\mathfrak{q} A_{2}}{P_{0}\mathcal{P}_{2}}\equiv \frac{J_{1}A_{2}}{P_{0}\mathcal{P}_{0}}, \label{velocityind}\end{align}

\vspace{-0.1cm}
\noi where it has been used the relation Eq.\ (\ref{condition}) with $J_{2}=0$, to express Eq.\ (\ref{velocityind}) in two different ways. It follows that in this particular configuration the two BH horizons read
\vspace{-0.2cm}
\begin{align}
\sigma_{1}&=\sqrt{D_{1}-\mathfrak{q}^{2}\mathcal{P}_{0}\left( \frac{\mathcal{P}_{0}G_{1}-2A_{1}B_{1}\mathcal{P}_{2}}{P_{0}^{2}\mathcal{P}_{1}^{2}\mathcal{P}_{2}^{2}}\right)},\nonu\\
\sigma_{2}&=\sqrt{D_{2}}.\label{sigmasII}\end{align}

\vspace{-0.1cm}
\noi where $D_{i}$ is defined in Eq.\ (\ref{sigmas}). It is quite evident how the first BH induces rotation to the second one via the parameter $\mathfrak{q}$, where it should be observed that the second BH horizon does not remain purely electrostatic [see Eq.\ (\ref{VarzChist})] due to its interaction with the first rotating BH. Somehow the induced rotation on the second BH is hidden by its own electric charge. A similar description for the induction of electric charge can be added too, if we suppose now that $Q_{2}=0$, after considering a careful calculation, the non zero electric potential $\Phi_{2}^{H}$ acquires the simple aspect
\vspace{-0.2cm}
\begin{align} \Phi_{2}^{H}&=\frac{Q_{1}(R+M)}{P_{0}}, \label{inducedpotential}\end{align}

\vspace{-0.1cm}
\noi and the induction of electric charge from the first BH to the second one is restricted to solve for $\mathfrak{q}$ the next cubic equation
\vspace{-0.1cm}
\begin{align} M_{1}\big(\mathfrak{q}P_{0}-a_{1}p_{1}-a_{2}p_{2}\big)\nonu\\
-Q_{1}^{2}\big[M_{2}\mathfrak{q}+(M_{1}-M_{2})a_{2}\big]=0, \label{relacionnocharge}\end{align}

\vspace{-0.1cm}
\noi which arises directly from Eq.\ (\ref{condition}) once is fixed $Q_{2}=0$. Thus having the following BH horizons
\vspace{-0.1cm}
\begin{widetext}
\begin{align}
\sigma_{1}&=\sqrt{M_{i}^{2}- Q_{1}^{2}\left(\frac{p_{1}}{P_{0}}+\frac{Q_{1}^{2}\mathfrak{q}^{2}a_{0}^{2}}{P_{0}^{2}c_{0}^{2}}\right)
-J_{1}\left(\frac{J_{1}g_{1}-2\mathfrak{q}a_{0}b_{1}}{P_{0}^{2}c_{0}^{2}}\right)},\quad
\sigma_{2}=\sqrt{M_{2}^{2}-a_{2}^{2}\left(1+\frac{Q_{1}^{2}\big(2p_{2}+Q_{1}^{2}\big)}{P_{0}^{2}}\right)+\gamma_{21}},\nonu\\
g_{1}&=p_{1}\Big[p_{1}\big(p_{2}+Q_{1}^{2}\big)^{2}-4M_{2}P_{0}c_{0}\Big], \quad a_{0}=c_{0}-M_{1}\big(R^{2}-\Delta_{1}\big),\quad
b_{1}= Q_{1}^{2}\big(R^{2}-\Delta_{1}\big)p_{2}-2c_{0}h_{1}, \nonu\\
c_{0}&=M_{1}p_{2}+(M_{1}-M_{2})Q_{1}^{2}, \quad
h_{1}= M_{1}P_{0}-Q_{1}^{2}(R+M),\quad \Delta_{1}=M^{2}-Q_{1}^{2}-\mathfrak{q}^{2},
\label{sigmaselectro}
\end{align}
\end{widetext}

\vspace{-0.1cm}
Another physical property of these configurations is the interaction force related to the conical singularity, which is understood as a measure of how much pressure each source is feeling during the dynamical interaction, where the strut prevents both sources from collapsing over each other. In order to calculate the force we use the formula $\mathcal{F}=(e^{-\gamma_{s}}-1)/4$ \cite{Israel,Weinstein}, where $\gamma_{s}$ defines the constant value that metric function $\gamma$ takes on the region of the strut. The force in the electrovacuum case, after a lengthy algebraic procedure with the aid of Eqs.\ (\ref{condition}) and (\ref{sigmas}) assumes the final form
\vspace{-0.1cm}
\begin{align}
\mathcal{F}&=\frac{\mathcal{N}_{0}}{P_{0}^{3}(R^{2}-M^{2}+Q^{2}+\mathfrak{q}^{2})},  \nonu\\
\mathcal{N}_{0}&=(M_{1}M_{2}P_{0}^{2}-\mathfrak{q}^{2}Q_{1}^{2}Q_{2}^{2})\left[(R+M)^{2}
-\mathfrak{q}^{2}\right]\nonu\\
&-(Q_{1}-F_{0})(Q_{2}+F_{0})P_{0}^{3}
+\mathfrak{q}^{2}\Big\{(M_{1}Q_{2}-M_{2}Q_{1})^{2}P_{0}\nonu\\
+&Q_{1}Q_{2}\left[ 2(R^{2}+MR+\mathfrak{q}^{2})P_{0}+(P_{0}+Q_{1}Q_{2})Q^{2}\right] \Big\}.
\label{force} \end{align}

\vspace{-0.1cm}
The expression of the force offers relevant information to define the limits of the interaction distance between the BHs. For instance, the minimal distance at which both horizons are touching each other (the ML) is reached when the force tends to its maximum value; i.e., $\mathcal{F} \rightarrow \infty$, and this is possible only if the denominator is equal to zero. Hence, one may conclude that the distance $R\equiv R_{0}=\sqrt{M^{2}-Q^{2}-\mathfrak{q}^{2}}= \sigma_{1}+\sigma_{2}$ defines the ML, while on the other hand, from Eq.\ (\ref{conditionparameters}) one gets the value $\mathfrak{q}=J/M$. Surprisingly, in the ML it can be possible to obtain very simple formulas for $\Omega_{i}$,
$\Phi_{i}^{H}$, $\kappa_{i}$, and $S_{i}$, which are given by
\vspace{-0.1cm}
\begin{align}
\Omega_{i}&=\frac{J/M}{d_{0}},\qquad \Phi_{i}^{H}=\frac{Q(R_{0}+M)}{d_{0}}, \nonu\\ \frac{S_{i}}{4\pi}&=\frac{\sigma_{i}}{\kappa_{i}}=\frac{d_{0}\sigma_{i}}{R_{0}},\qquad
d_{0}=\big(R_{0}+M\big)^{2}+\big(J/M\big)^{2},\nonu\\
R_{0}&=\sqrt{M^{2}-Q^{2}-(J/M)^{2}},
\label{HorizonpropertiesII}\end{align}

\vspace{-0.1cm}
\noi whereas the final values of the BH horizons at this limit are eventually simplified as
\vspace{-0.2cm}
\be \sigma_{i}=M_{i}-\frac{Q_{i}Q(R_{0}+M)+2J_{i}(J/M)}{d_{0}}. \label{Smarrfusion}\ee

\vspace{-0.1cm}
\noi Notice that Eq.\ (\ref{Smarrfusion}) is nothing less than the Smarr formula for the mass, which still holds in the ML. As a matter of fact, both BHs acquire the same final value on their thermodynamical properties at this limit. The extreme case for BHs $(\sigma_{i}=0)$ during the ML can be also considered here; it emerges after setting $R_{0}=0$ in Eq.\ (\ref{Smarrfusion}), where both angular momenta $J_{i}$ satisfying such a condition have the form
\vspace{-0.1cm}
\begin{align} J_{i}&=M_{1}\sqrt{M^{2}-Q^{2}}-\frac{Q(M_{2}Q_{1}-M_{1}Q_{2})}
{2\sqrt{M^{2}-Q^{2}}}, \nonu \\
J_{2}&=M_{2}\sqrt{M^{2}-Q^{2}}+\frac{Q(M_{2}Q_{1}-M_{1}Q_{2})}
{2\sqrt{M^{2}-Q^{2}}},\end{align}

\vspace{-0.1cm}
\noi and it follows that the sum is recovering the well-known expression for extreme KN BHs, namely
\vspace{-0.1cm}
\be J_{1}+J_{2}=(M_{1}+M_{2})\sqrt{(M_{1}+M_{2})^{2}-(Q_{1}+Q_{2})^{2}}. \label{extremerelation} \ee

\vspace{-0.1cm}
Thus, we have that the merger process is producing one single extreme KN BH of mass $M=M_{1}+M_{2}$, angular momentum $J=J_{1}+J_{2}$, and total electric charge $Q=Q_{1}+Q_{2}$. On the other hand, with the aim to gain more insight on how the sources are affecting to each other at large distances it is necessary to observe the asymptotic behavior of the interaction force when $R \rightarrow \infty$, therefore, we use Eqs.\ (\ref{condition}) and (\ref{force}) to obtain
\vspace{-0.1cm}
\begin{align} \mathcal{F}&\simeq\frac{M_{1}M_{2}-Q_{1}Q_{2}}{R^{2}} \Bigg[1+\frac{M^{2}
-Q^{2}-3(a_{1}+a_{2})^{2}}{R^{2}} \nonu\\
&-\frac{M_{2}Q_{1}-M_{1}Q_{2}}{M_{1}M_{2}-Q_{1}Q_{2}}\Bigg(\frac{Q_{1}
-Q_{2}}{R} \nonu\\
&-\frac{2(M_{2}Q_{1}-M_{1}Q_{2})+M_{1}Q_{1}-M_{2}Q_{2}}{R^{2}}\Bigg)\nonu\\
&+ O \left(\frac{1}{R^{3}} \right)\Bigg],\label{asymptoticforce} \end{align}

\vspace{-0.1cm}
\noi and since $\mathfrak{q} \rightarrow a_{1}+a_{2}$ when the sources are far away from each other, one may recover from Eq.\ (\ref{sigmas}) the formula $\sigma_{i}=\sqrt{M_{i}^{2}-Q_{i}^{2}-J_{i}^{2}/M_{i}^{2}}$ representing a single KN BH. It is pretty much clear that expression Eq.\ (\ref{asymptoticforce}) extends the formula of Dietz and Hoenselaers \cite{DH} that describes the spin-spin interaction at large distances, which is recovered after killing the contribution of both electric charges.

Finally, we are going to discuss the physical scenario in which the strut is removed, where is crucial to impose the condition $\mathcal{F}=0$. In this circumstance the numerator of $\mathcal{F}$ defines a bicubic equation in terms of the variable $\mathfrak{q}$ that assumes the form
\vspace{-0.1cm}
\begin{align}
&\mathfrak{q}^{6}+3A_{1}\mathfrak{q}^{4}+3A_{2}\mathfrak{q}^{2}+A_{3}=0,\nonu\\
A_{1}&= \frac{(R+M)^{2}\mu_{+}-Q_{1}Q_{2}\Big[2R(R+M)+Q^{2}+Q_{1}Q_{2}\Big]}{3(M_{1}M_{2}-Q_{1}Q_{2})}, \nonu\\
A_{2}&=(R+M)^{2}\bigg(A_{1}-\frac{(R+M)^{2}}{3}\bigg)\nonu\\
&-\frac{(R+M)^{2}\Big((R+M)^{2}\mu_{-}-2Q_{1}^{2}Q_{2}^{2}\Big)+Q_{1}^{2}Q_{2}^{2}Q^{2}}
{3(M_{1}M_{2}-Q_{1}Q_{2})}, \nonu\\
A_{3}&=-\frac{(R+M)^{6}\mu_{-}}{M_{1}M_{2}-Q_{1}Q_{2}},  \nonu\\
\mu_{\pm}&= M_{1}M_{2}\pm \bigg(Q_{1}-\frac{M_{2}Q_{1}-M_{1}Q_{2}}{R+M_{1}+M_{2}}\bigg) \nonu\\
&\times \bigg(Q_{2}+ \frac{M_{2}Q_{1}-M_{1}Q_{2}}{R+M_{1}+M_{2}}\bigg),
\label{nostrut} \end{align}

\vspace{-0.1cm}
\noi where fortunately this bicubic equation can be solved analytically. Before giving its explicit solution, let us show that if any of the electric charges is set to zero, for instance $Q_{2}=0$, this equation together with Eq.\ (\ref{condition}) satisfy the following result
\vspace{-0.1cm}
\begin{align}
&J_{1}+J_{2}+R\bigg(\frac{J_{1}}{M_{1}}+\frac{J_{2}}{M_{2}}\bigg)-(R+M_{1}+M_{2})\mathfrak{q}=0, \nonu\\
\mathfrak{q}&=-\epsilon\sqrt{(R+M_{1}+M_{2})^{2}-\frac{Q_{1}^{2}(R+M_{2})}{M_{1}}}, \nonu\\
\epsilon &= \pm 1, \label{nostrutI}
\end{align}

\vspace{-0.1cm}
\noi and $Q_{1}=0$ leads to the equilibrium law for vacuum systems \cite{MankoRuiz}. The case concerning unequal counterrotating KN binary BHs \cite{ICM} that is obtainable from Eqs.\ (\ref{condition}) and (\ref{nostrut}) by fixing $\mathfrak{q}=0$, is another special case, which permits an explicit relation between the physical parameters, namely
\vspace{-0.1cm}
\begin{align}
M_{1}M_{2}&- \bigg(Q_{1}-\frac{M_{2}Q_{1}-M_{1}Q_{2}}{R+M_{1}+M_{2}}\bigg) \nonu\\
&\times \bigg(Q_{2}+ \frac{M_{2}Q_{1}-M_{1}Q_{2}}{R+M_{1}+M_{2}}\bigg)=0, \nonu\\
&J_{1}+J_{2}+R\bigg(\frac{J_{1}}{M_{1}}+\frac{J_{2}}{M_{2}}\bigg)=0,
\label{nostrutII}
\end{align}

\vspace{-0.1cm}
\noi where in the lack of rotation it is reduced to the static relation first discovered in \cite{Alekseev}. Nevertheless, the most general case that solves Eq.\ (\ref{nostrut}), and gives the chance to balance the sources once the strut has been eliminated turns out to be
\vspace{-0.1cm}
\begin{align}
\mathfrak{q}^{2}_{(k)}&= -A_{1} + e^{i 2\pi k/3}\big[\mathfrak{b}_{o}+ \sqrt{\mathfrak{b}_{o}^{2}-\mathfrak{a}_{o}^{3}}\big]^{1/3} \nonu\\
&+ e^{-i 2\pi k/3}\mathfrak{a}_{o} \big[\mathfrak{b}_{o} + \sqrt{\mathfrak{b}_{o}^{2}-\mathfrak{a}_{o}^{3}}\big]^{-1/3},  \quad
\mathfrak{a}_{o}=A_{1}^{2}-A_{2},\nonu\\ 
\mathfrak{b}_{o}&=(1/2)\big[3A_{1}A_{2}-A_{3}-2A_{1}^{3}\big], \quad k=0,1,2. \label{theq} \end{align}

\vspace{-0.1cm}
\begin{figure}[ht]
\begin{minipage}{0.49\linewidth}
\centering
\includegraphics[width=4.3cm,height=4.5cm]{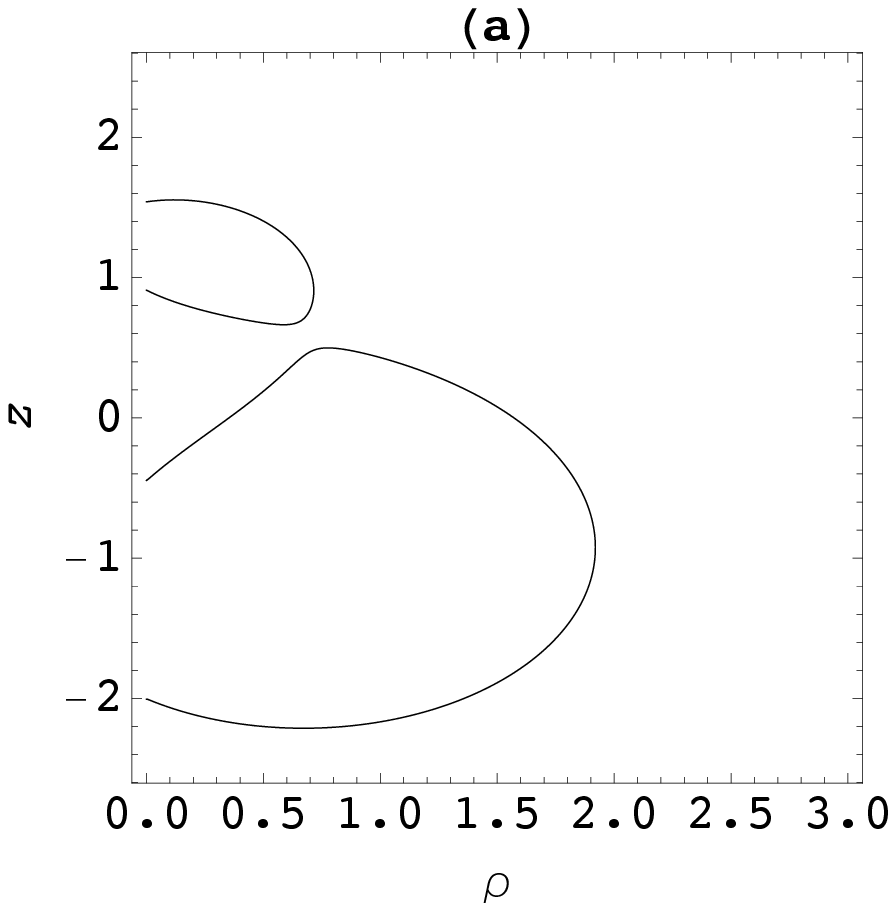}
\end{minipage}
\begin{minipage}{0.49\linewidth}
\centering
\includegraphics[width=4.3cm,height=4.5cm]{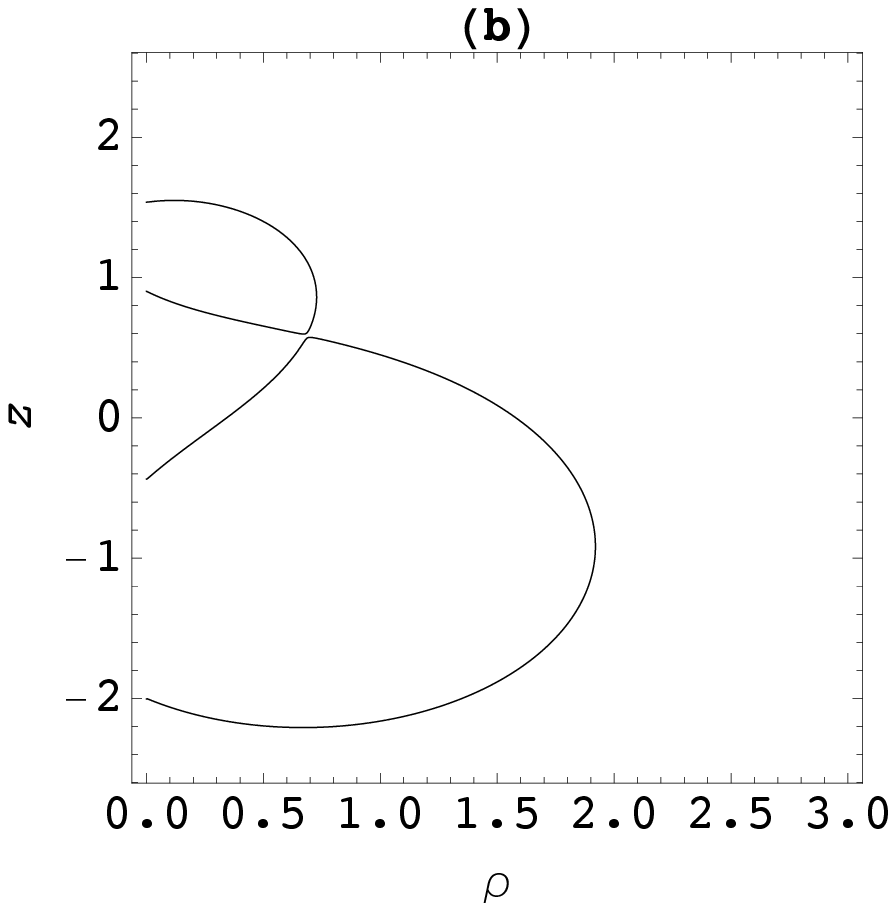}
\end{minipage}\vspace{0.1cm}
\begin{minipage}{0.49\linewidth}
\centering
\includegraphics[width=4.3cm,height=4.5cm]{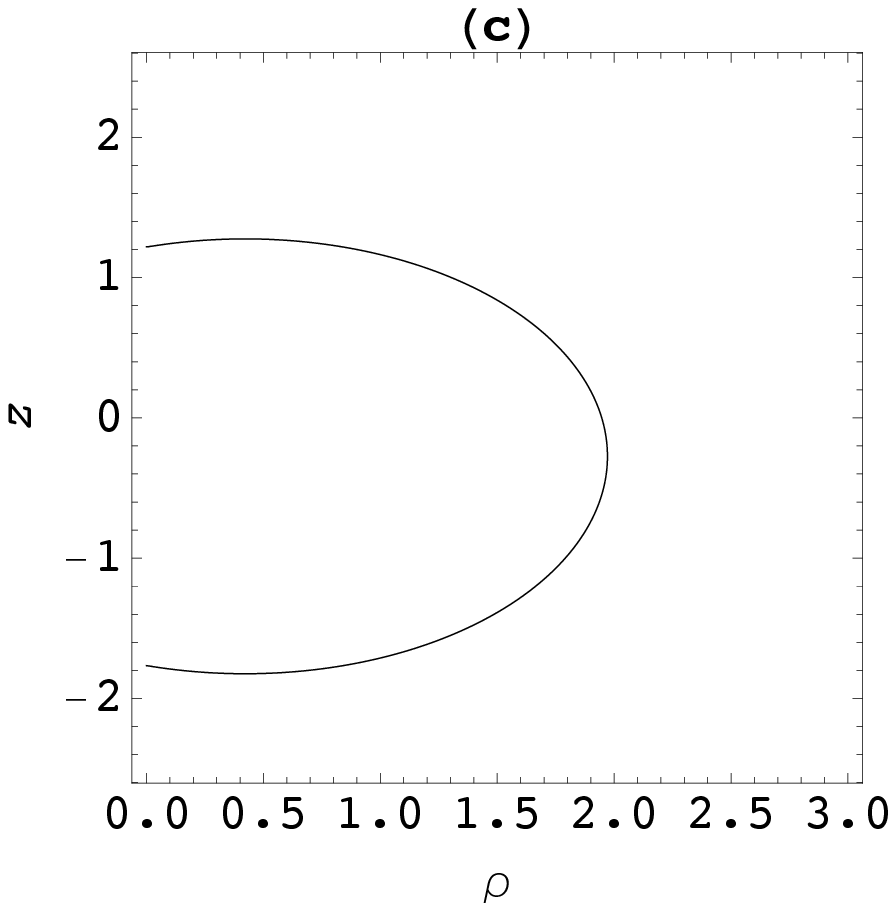}
\end{minipage}
\begin{minipage}{0.49\linewidth}
\centering
\includegraphics[width=4.3cm,height=4.5cm]{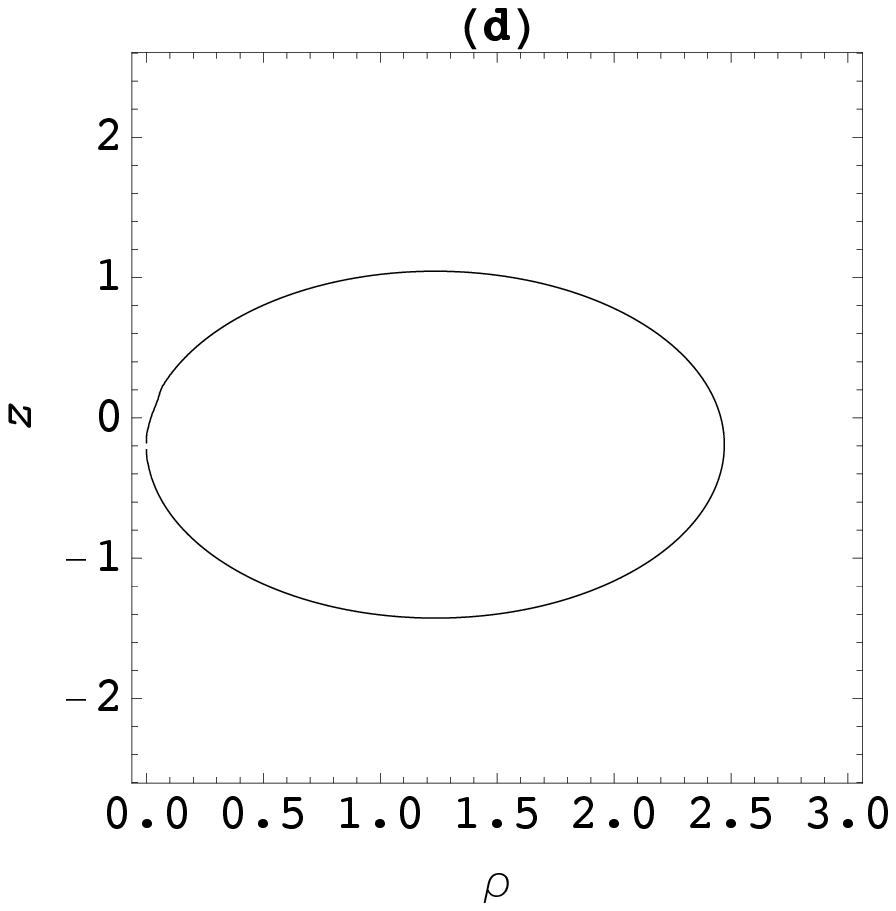}
\end{minipage}
\vspace{-0.1cm}
\caption{The stationary limit surfaces $(f=0)$ of the BS in the presence of the strut for the values $M=1$, $M=2$, $a_{1}=0.9$, $a_{2}=2.507$, $Q_{1}=1.2$, $Q_{2}=0.5$, and various distances $R$; (a) $R=2.45$, (b) $R=2.439$, (c) $R=1.4913$ (the ML). (d) During the merger process at $R=0.0047$, the first BH tends to increase its rotation acquiring the value $a_{1}=2.4$ in order to compensate the repulsion of the second one, establishing an equilibrium state with no strut in between BHs, where the new source agrees with the condition $M^{2}-Q^{2}-(J/M)^{2}\geq 0$.}
\label{MPI}\end{figure}

\vspace{-0.2cm}
\begin{table}[ht]
\begin{tabular}{c c c c c c c}
\hline \hline
$\sigma_{1}$&$\sigma_{2}$&$ a_{1}$ & $a_{2}$ &  $Q_{1}$& $R$ & $\mathfrak{q}$ \\ \hline
0.1872 & 0.2023 & 2.4 & 2.507 & 1.2   & 0.0047   & 2.4756  \\
0.1815 & 0.2321 & 1.1 & -2.79 & 2.1   & 0.0310   & -1.4949  \\
0.7914 & 0.3917 & 0.7 & 4.180 & -0.2   & 0.0896   & 3.0727  \\
1.2300 & 0.9676 & 0.7 &-4.890 & -0.2   & 0.0685   &-3.0520  \\
0.4265 & 0.1316 & 1.2 & 3.476 & 0.8   & 0.0567   & 2.7571  \\
0.9666 & 0.2492 & -0.9 & 4.8 & -1.5   & 0.1557   & 2.9416  \\
  \hline \hline
\end{tabular}
\centering
\vspace{-0.2cm}
\caption{Equilibrium states fixing the values $M_{1}=1$, $M_{2}=2$, and $Q_{2}=0.5$.  }
\label{table1}
\end{table}

\begin{figure}[ht]
\begin{minipage}{0.49\linewidth}
\centering
\includegraphics[width=4.3cm,height=4.5cm]{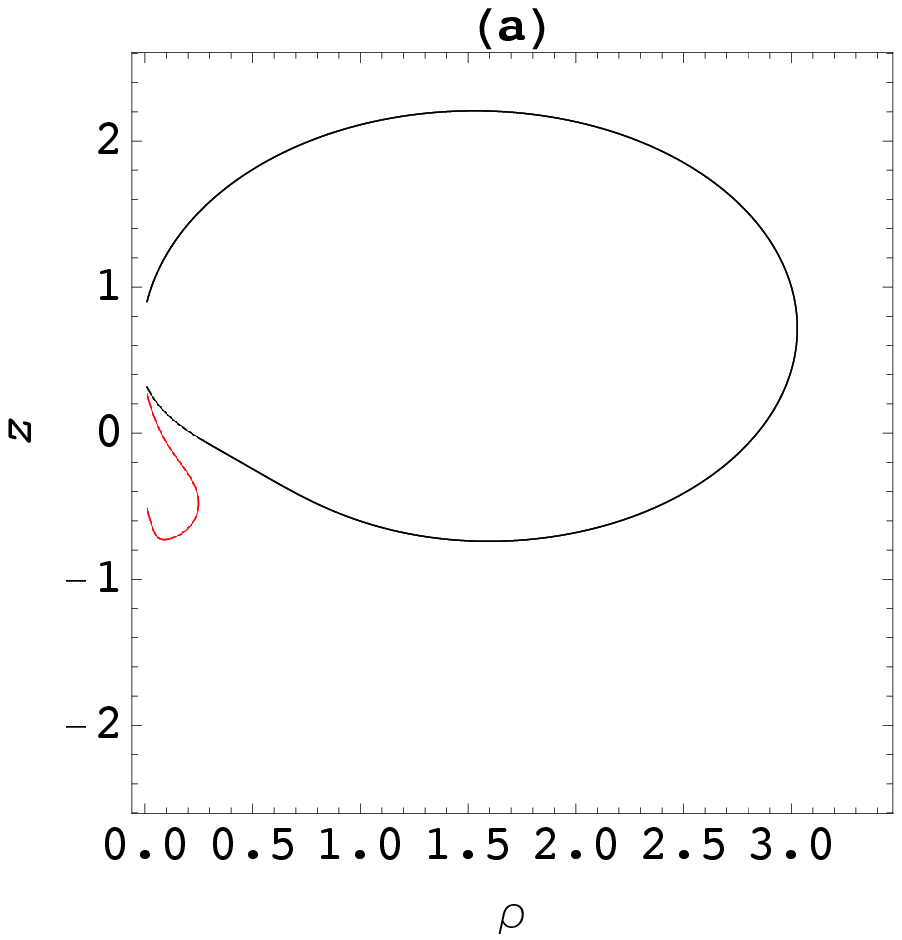}
\end{minipage}
\begin{minipage}{0.49\linewidth}
\centering
\includegraphics[width=4.3cm,height=4.5cm]{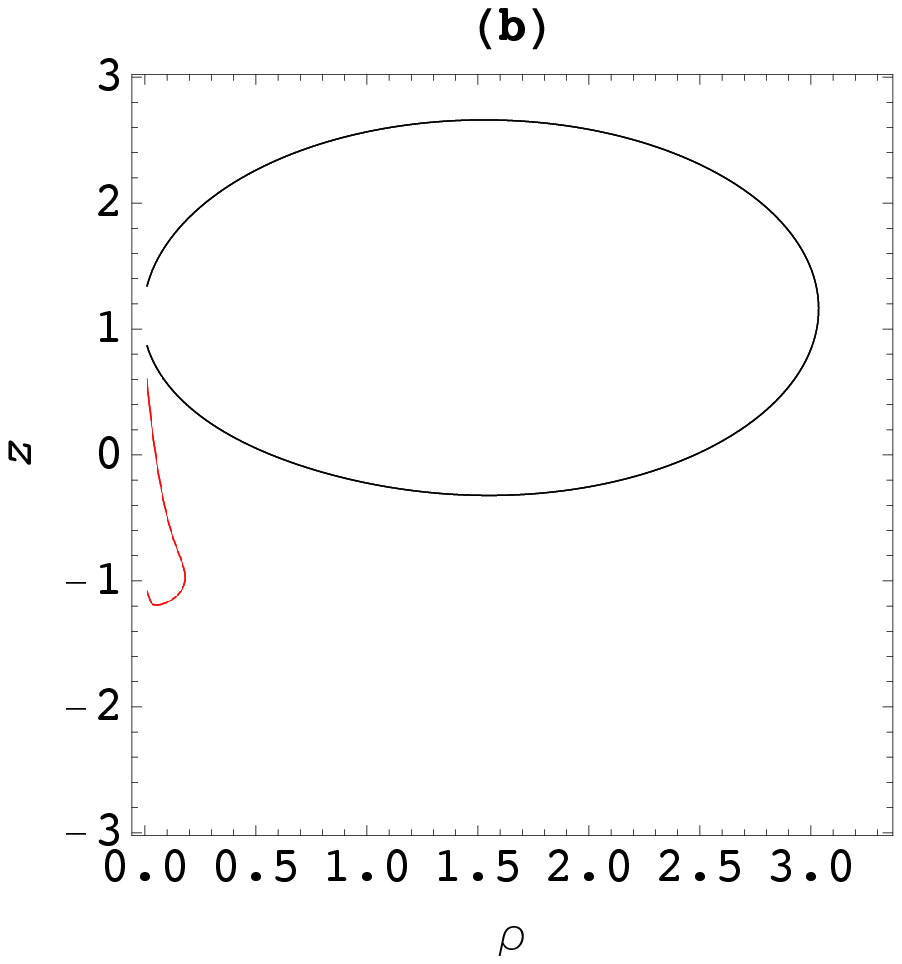}
\end{minipage}
\vspace{-0.1cm}
\caption{Appearance of CTC during the merger process in the absence of a strut, for the values $M=1$, $M=2$, $Q_{1}=-0.2$, $Q_{2}=0.5$, $a_{1}=0.7$ and (a) $a_{2}=4.180$, $R=0.0896$, (b) $a_{2}=-4.890$, $R=0.0685$. The condition $M^{2}-Q^{2}-(J/M)^{2}<0$ holds in both cases.}
\label{MPII}\end{figure}

\vspace{-0.1cm}
In the search for equilibrium configurations in which the condition $\mathcal{F}=0$ is satisfied, similarly to the vacuum scheme \cite{Cabrera2018}, we have not been able to find equilibrium states before the BHs reach the merger limit (at the distance $R_{0}=\sqrt{M^{2}-Q^{2}-(J/M)^{2}}$), except during the merger process. This laborious task has been done for a wide range of numerical values. In Table \ref{table1} are displayed several numerical values fulfilling equilibrium states during the merger process where the first and second sets of values define a source that apparently can be seen as a BH since it satisfies the condition $M^{2}-Q^{2}-(J/M)^{2}\geq 0$ [see Fig.\ \ref{MPI}(d)]. Moreover, in the remaining sets of values, such a condition is negative and therefore the newly formed source cannot be considered a BH due to the appearance of closed timelike curves (CTC). The CTC are contained within the region defined by $g_{\varphi \varphi} \equiv \rho^{2}f^{-1}-f\omega^{2}<0$ as shown in Fig.\ \ref{MPII}. This result agrees with the description that the CTC region of a KN BH must be always contained inside the event
horizon \cite{Bonnor}.

\vspace{-0.1cm}

\vspace{-0.3cm}
\section{Summary and outlook}
\vspace{-0.4cm}
In this paper we have reported the most general exact model that permits a description of two unequal interacting KN BHs separated by a massless strut in stationary axisymmetric spacetimes. First of all, the conditions on the axis and the one eliminating both magnetic charges were established and solved analytically through an appropriate parametrization that eventually allowed us the derivation of non-trivial formulas for the half-length horizons $\sigma_{i}$, as a function of seven arbitrary physical parameters of the system $\{M_{i},J_{i},Q_{i},R\}$, $i=1,2$.
As a matter of fact, such formulas are generalizations of the vacuum problem \cite{Cabrera2018} and the electrostatic one \cite{VCH}, where also other simpler rotating models endowed with electric charges can be trivially found \cite{CCHV,ICM}. Subsequently, the whole thermodynamical characteristics of the BS have been also depicted by very concise expressions. It is noteworthy that in the same manner as in the vacuum scenario \cite{Cabrera2018}, there exists a dynamical law for interacting KN BHs with struts, which is now defined by a septic algebraic equation. Unfortunately, there is no chance to solve analytically this higher degree equation, however, it has been quite helpful in better understanding the physical limits of the solution. The well-known equilibrium cases with no strut \cite{MankoRuiz,Alekseev} can be directly derived from this higher degree equation. On the other hand, the strut can be eliminated during the merging process, where unlike the vacuum case, apparently the new source that has been created is a BH, which is also free of some pathologies like CTC. We believe that this novel result opens new expectations in further studies with an astrophysical meaning. For instance, it would be plausible to consider 
the study of the induction of rotation or charge from one BH to another, within the framework of the well-known Penrose energy extraction process \cite{Penrose}. 

The extension of the binary model to include unequal magnetic charges on each KN BH, and thus define a rotating dyonic binary BH system it might be possible to achieve once a duality rotation procedure a la Carter \cite{Carter} is applied \cite{Cabrera2020}. However, even though it seems quite trivial to achieve this goal when electrical charges are identical, in the unequal case this problem can be hard to deal with. We expect to consider this and other issues related to interacting binary BHs in the future.

\vspace{-0.6cm}
\section*{Acknowledgements}
\vspace{-0.4cm}
The author acknowledges the financial support of SNI-CONACyT, M\'exico, grant with CVU No. 173252.

\end{document}